\documentclass{elsart}

\usepackage{natbib}

\usepackage{amssymb}

\begin{document}
\def\be{\begin{equation}}
\def\ee{\end{equation}}
\def\bc{\begin{center}} 
\def\ec{\end{center}}
\def\bea{\begin{eqnarray}}
\def\eea{\end{eqnarray}}

\begin{frontmatter}

 \title{Number of cliques in random scale-free network  ensembles}

 \author{Ginestra Bianconi}
 \ead{gbiancon@ictp.it}
 
 \address{the Abdus Salam ICTP, Strada Costiera 11, 34014 Trieste,Italy}

\author{Matteo Marsili}
 \address{the Abdus Salam ICTP, Strada Costiera 11, 34014 Trieste,Italy} 
\ead{marsili@ictp.it}

\begin{abstract}

In this paper we calculate the average number of cliques in random scale-free networks. We consider first the hidden variable ensemble and subsequently the Molloy Reed ensemble. In both cases  we find  that cliques, i.e. fully connected subgraphs,  appear also when the average degree is finite. This is in contrast to what happens  in Erd\"os and Renyi  graphs in which diverging average degree is required to observe cliques of size $c>3$. Moreover we show that in random scale-free networks the clique number, i.e. the size of the largest clique present in the network diverges with the system size. 
\end{abstract}


\end{frontmatter}

When graphs are  used to represent a variety of real  technological, social and biological  systems they are  called networks.
The analysis of many real networks reveals that while different networks differ one  from another in their local structure,  characterized by operational modules or {\em motifs} that are a signature of their function \citep{Milo,Vazquez_m,Dobrin}, many networks have some important common characteristics \citep{RMP,Doro,Internet}.
In particular a  large variety of networks have been shown to display a scale-free degree distribution $P(k)\sim k^{-\gamma}$ with non universal $\gamma$ exponents.
 The scale-free degree distribution strongly affects the local topology of the networks. For example,  scale-free networks with an exponent $\gamma<3$ have a very large number of small loops \citep{Loops,lungo}, which is a very distinctive feature with respect to Erd\"os and Renyi (ER)  networks with finite average connectivity \cite{Janson,Monasson}.  In its turn, this very peculiar local structure induce  many relevant effects of the dynamics defined on these networks \citep{Ising1,Ising2,Havlin, Vespi}.

A special type of network subgraphs are cliques, i.e.  fully connected  subsets of nodes of the network. Cliques are  relevant  objects for the study of  real networks, in fact cliques and overlapping sets of cliques  provide relevant insights on the community  structure of  networks \citep{Vicsek1,Vicsek2}. In random Erd\"os and Renyi graphs of $N$ nodes and  linking probability $p(N)$ the expected number of cliques \citep{Janson} of size $c$ is given by
\be
\langle { N}_c^{ER}\rangle=\left(\begin{array}{l} N\nonumber\\
c\end{array}\right)p(N)^{c(c-1)/2}.
\ee
Consequently in the large $N$ limit  the expected number of small cliques with $c>3$ is different from zero only when the average degree $\langle k\rangle$
diverges as $N\rightarrow \infty$. 
Special attention in mathematics literature is  given to the  maximal size of the clique present in a graph $G$, i.e. its clique number $c_{max}$.
The clique number is an important characteristic of networks and constitute also a lower bound to the coloring number, since in the coloring problem one is forced to color all the nodes of a clique with a different color. In \cite{clique1}. we show that scale-free networks with $\gamma<3$ have many more  and larger cliques than random Erd\"os and Renyi networks. 

In this paper we provide the complete derivation of the  theoretical expectation on the number of cliques in random scale-free networks. We do this by evaluating the average number of cliques and its second moment.
We found the surprising result that cliques of size $c>3$  are present also in networks with finite average connectivity, i.e. networks with $\gamma\in(2,3]$.
Moreover we  can prove that the clique number $c_{max}$ of networks with $\gamma<3$ diverge with the network size $N$ providing  upper and lower bounds for the clique number.
These bound arise from classical inequalities for probabilities which involve the first and the second moment of the number of cliques. These can be computed in different ensembles of random graphs  \citep{MR,Kahng}.
The main section of this paper would be devoted to the calculation of the average number of cliques and its second moment in the hidden variable ensemble \citep{HV1,HV2}.
Subsequently the derivation of the average number of clique is extended to the Molloy Reed ensemble \cite{MR}. The same scaling of the number of cliques is found also in this ensemble.
Finally the conclusions are given.

\section{Hidden variable ensemble} 

In this ensemble the prescription  to generate a class of  scale-free networks with exponent $\gamma$ is the following: {\em i)} assign to each node $i$ of the graph a hidden continuous variable $q_i$ distributed according a $\rho(q)$ distribution. Then {\em ii)} each pair of nodes with hidden variables $q,q'$ are linked with probability $r(q,q')$.
When the hidden variable distribution is scale-free $\rho(q)=\rho_0 q^{-\gamma}$ for $q\in [m,Q]$  and  the linking probability is linear in both $q$ and $q'$,i.e. $r(q,q')=qq'/(\langle q\rangle N)$ we obtain a random uncorrelated scale-free network.
In this specific case  a cutoff    
\bea
Q\sim \left\{\begin{array}{lcr}
N^{1/\gamma}& \mbox{for} &\gamma \in(1,2] \nonumber \\
N^{1/2}& \mbox{for}&\gamma\in(2,3]  \nonumber \\
N^{1/(\gamma-1)}&\mbox{for}&\gamma\in (3,\infty) \nonumber \\
\end{array}\right.
\eea
is needed to keep the linking probability smaller than one, i.e. $Q^2/(\langle q\rangle N)\leq 1$.

\subsection{ Average number of cliques}
A clique $C$ of size $c$ is a set of $c$ distinct nodes $C = \{i_1,\ldots,i_c\}$, each one connected with all the others. For each choice of the nodes, the
probability that they are connected in a clique is
\be
\prod_{i\neq j\in C} r(q_{i},q_{j}).
\ee
Consequently the average number of cliques of size $c$ is  given by  the number of ways in which we can pick $c$ nodes in the network with $n(q)$ nodes with hidden variable $q_i\in (q,q+\Delta q)$ multiplied by the probability that each couple of node of this set is linked. Since in random scale-free networks we have  $r(q,q')=\frac{q q'}{\langle q\rangle N}$  we can write
\bea
\langle{N}_c \rangle &=&\sum_{\{n(q)\}}' \prod_{q} \left(\begin{array}{l} N(q)\\ n(q) \end{array}
\right) \left(\frac{q}{\sqrt{\langle q\rangle N}}\right)^{(c-1)n(q)}\eea
where   $N(q)=N\rho(q)$ are the  nodes of the network with hidden variable  $q_i\in (q,q+\Delta q)$ and where the sum is extended to all the sequences  $\{n(q)\}$  satisfying $\sum_q n(q)=c$. Introducing a integral representation of the delta function $\delta(\sum_q n(q)-c)$  and performing the summation over $n(q)$ we get
\bea
\langle{N}_c \rangle = \int dy e^{y c+ N\langle\log [1+{\theta}^{c-1}e^{-y}]\rangle}.
\label{N_m}
\eea
where  we have taken the  limit $\Delta q\rightarrow 0$. In $(\ref{N_m})$ we have introduced   the variable $\theta$  defined as 
\be
{\theta}=\frac{q}{\sqrt{\langle q \rangle N }},
\label{calq}
\ee 
and we have indicated with  $\langle \rangle$  the average over the distribution $\rho(q)$.
Solving the integral in $(\ref{N_m})$ by saddle point method one finds
\bea
\langle {N}_c\rangle \simeq \sqrt{\frac{2\pi}{N|f''(y^*)|}}  e^{Nf(y^*)}
\label{N_msp}
\eea
with $f(y)=yc/N+\langle\log [1+{\theta}^{c-1}e^{-y}]\rangle $ and $y^*$ fixed by the saddle point equation 
\be
\frac{c}{N}=\left\langle\frac{{\theta}^{c-1}e^{-y^*}}{1+{\theta}^{c-1}e^{-y^*}}\right\rangle.
\label{ystar}
\ee
If we assume that the cutoff of the hidden variable distribution is equal to  $Q=\sqrt{\langle q \rangle N}(1-\epsilon)$ with an $\epsilon\geq 0$, the maximal clique size depends on both the $\gamma$ exponent and on $\epsilon$.
The dependence in $\epsilon$ reflects the fact that when $\epsilon=0$ the highest degree nodes have a probability to be linked $r(q,q')$ which approach one.
Considering the  definition of $y^*$ from  Eq. $(\ref{ystar})$ we can see that
the asymptotic expansion
\be
e^{y^*}\approx \frac{N}{c}\langle{\theta}^{c-1} \rangle 
\ee
is valid until 
\be
\frac{c}{N}
\frac{\left\langle{\theta}^{2(c-1)}\right \rangle }{
\left[\left\langle{\theta}^{(c-1)}\right \rangle \right]^2}\leq Q^{\gamma-1}\frac{c^2}{(2c-3)N} < 1,
\label{bounds.eq}
\ee
i.e. until $c< c^*\sim(2 N Q^{1-\gamma})^{1/2}$.

Consequently for  clique sizes $c<c^*$ one has the valid asymptotic expression for $\langle {N}_c\rangle$
\be
\langle {N}_c\rangle\simeq \sqrt{\frac{2\pi}{c}} \left[\left\langle\frac{Ne}{c}{\theta}^{c-1}\right\rangle\right]^c.
\label{asym}
\ee

To find an upper bound for  the clique number (the maximal clique size) is a bit more involved.
We start from the classical inequality
\be
P({N}_c>0)\leq \langle {N}_c\rangle 
\label{undici}
\ee
and the expression $(\ref{N_msp})$ together with $(\ref{ystar})$ for the average number of cliques $\langle {N}_c\rangle$. If $ \langle {N}_{\bar{c}}\rangle\rightarrow 0$ in the $N\rightarrow \infty$ limit, then $\bar{c}$  fixes an upper bound $\bar{c}$ for the maximal clique size of the network. 

In Appendix A  show that  the  clique number  $c_{max}\leq \bar{c}$ with $\bar{c}$ satisfy for $\epsilon=0$ the condition
\be
\left\langle\frac{N e^2}{\bar{c}}{\theta}^{\bar{c}-1}\right\rangle=1.
\label{cbar}
\ee
On the other this expression provide a upper bound also for the case $\epsilon\neq 0$ since in this case $\bar{c}$ defined in Eq. $(\ref{cbar})$ is still in within the validity of the asymptotic expansion $(\ref{asym})$ and correspond to an expected number of cliques $N_c\rightarrow 0$ as $N\rightarrow \infty$.
 
The values of $\bar{c}$ and $c^*$ will depend both on the $\gamma$ exponent and on the  value of $\epsilon$. In fact networks with different values of $\gamma$ have different structural cutoffs
\begin{itemize}
\item{\em Networks with $\gamma>3$}
\\
These networks have a natural cutoff $Q=a N^{\frac{1}{\gamma-1}}$.
Considering  this cutoff when performing  the average in  equation $(\ref{cbar})$, we find   $\bar{c}=3$ in the limit $N\rightarrow \infty$. Therefore these networks, as well as the Erd\"os and Renyi networks, have maximal clique size $c_{max}\leq 3$.
\item{\em Networks with $2<\gamma<3$} 
\\
These networks have a structural cutoff $Q=(1-\epsilon)\sqrt{\langle q\rangle N}$ and for $c<c^*$ the average number of cliques is given by
\bea
\langle {N}_c\rangle \approx \sqrt{\frac{2\pi}{c}}\left( A_{\gamma,\langle{q}\rangle}\frac{N^{(3-\gamma)/2} (1-\epsilon)^{(c-\gamma)} }{c(c-\gamma)}\right)^c.
\eea 
whit $A_{\gamma,\langle q \rangle}$ been a constant depending on the power-law exponent $\gamma $ and on the average connectivity of the graph $\langle q\rangle$.
Moreover the value of $\bar{c}$ and $c^*$ defined in equations $(\ref{bounds.eq})$ and $(\ref{cbar})$ depend on the system size $N$, the $\gamma$ exponent and on  $\epsilon$ as shown in the Table $\ref{table}$.

We observe that while for the case $\epsilon>0$ the  asymptotic expansion is valid much above the upper bound $\bar{c}$, for $\epsilon=0$ the upper bound and the limit of the validity of the asymptotic expansion $c^*$ have the same order of magnitude, i.e. $c^* \sim \bar{c}\sim N^{\frac{3-\gamma}{4}}$ but we have $\bar{c}>c^*$.

\item {\em Networks with $1<\gamma<2$}
\\
These networks have a structural cutoff  defined as in the case $2<\gamma<3$, i.e. $Q=(1-\epsilon)\sqrt{\langle q\rangle N }$.
Given this expression and the divergence of the average degree with the upper cutoff $\langle q\rangle\sim Q^{2-\gamma}$, we get that the upper cutoff $Q$ 
scales with the network size $N$ as
$Q\sim N^{1/\gamma}$.
The asymptotic expansion  gives for  the average number of cliques of sizes $c<c^*$ 
\bea
\langle {N}_c\rangle \approx \sqrt{\frac{2\pi}{c}} \left(B_{\gamma,m}\frac{N^{1/\gamma}e (1-\epsilon)^{(c+1-2/\gamma)} }{c(c-\gamma)}\right)^c
\eea
where $B_{\gamma,m}$ is a function depending on the power-law exponent $\gamma$ and on the lower cutoff $m$ of the distribution.

The value of $\bar{c}$ and $c^*$ defined in equations $(\ref{bounds.eq})$ and $(\ref{cbar})$ depend on the system size $N$, the $\gamma$ exponent and on  $\epsilon$. Their scaling is   shown in the Table $\ref{table}$.
Also in this range of values of $\gamma$ for  $\epsilon>0$ the  asymptotic expansion is valid much above the upper bound $\bar{c}$, while for  $\epsilon=0$ the upper bound and the limit of the validity of the asymptotic expansion $c^*$ have the same order of magnitude, i.e. $c^* \sim \bar{c}\sim N^{\frac{1}{2\gamma}}$, but we have $\bar{c}>c^*$.
\end{itemize}

\subsection{ Second moment of the average number of cliques}
In order to derive a lower bound on the clique number $c_{max}$ we use a classical relation of probability theory \citep{Janson},
i.e.
\be
P({N}_c>0)\ge {\langle{N}_c\rangle}^2/{\langle {N}^2_c\rangle}
\ee
 where $\langle {N}^2_c\rangle $ is the second moment of the number of cliques of size $c$ in the considered random graph ensemble. 
Consequently if ${{\langle{N}_c\rangle}^2}/{\langle {N}^2_c\rangle}\geq K $ we are guaranteed that the typical graph contains cliques of size $c$ with probability $P({N}_c>0)\geq K>0$. Thus we proceed in the calculation of the second moment of the clique number $\langle {N}^2_c\rangle$. To do this calculation we  count the average number of pairs of cliques of size $c$  present in the graph with an overlap of $o=0,\dots, c$ nodes. We use the notation $\{n(q)\}$ to indicate the number of the nodes with hidden variable $q$ belonging to the first  clique, $\{n_o(q)\}$ to indicate the number of nodes   belonging to the overlap and with $\{n'(q)\}$ to indicate  the number of nodes belonging to the second clique but not to the overlap. We consider only sequences $\{n(q)\}, \{n'(q)\}, \{n_o(q)\} $ which satisfy $\sum_q n(q)=c$, $\sum_q n_o(q)=o$ and  $\sum_q n'(q)=c-o$.
With these conditions, and then substituting the conditions with delta functions we get
\bea
\langle{N}^2_c \rangle&=&\sum_{o=0}^c \sum_{\{n(q)\}}'\sum_{\{n'(q)\}}'\sum_{\{n_o(q)\}}'  \prod_{q} \left(\begin{array}{c} N(q)\\ n(q) \end{array}
\right) \left(\begin{array}{c} N(q)-n(q)\\ n'(q) \end{array}
\right) \left(\begin{array}{c} n(q)\\ n_o(q) \end{array}
\right) {\theta}^{g(q)} \nonumber \\
&=&\int dy \int dy^{o} \int dy' \sum_{o=0}^c e^{N\langle f(y,y',y^o,q)\rangle}.
\eea
where $g(q)=(c-1)(n(q)+n'(q))+(c-o)n_o(q)$ and
\bea
f(y, y',y^o,q)&=&\frac{y c+y'(c-o)+y^{o}o}{N}+\nonumber \\
& & \log \left[1+\left(e^{-y'}+e^{-y}\right){\theta}^{c-1}+ e^{-(y+y^o)}{\theta}^{2c-o-1} \right].
\nonumber 
\eea
The saddle point method, gives
\bea
\left\{\begin{array}{l}
y=y'\nonumber \\
\frac{c-o}{N}=\left\langle \frac{e^{-y}{\theta}^{c-1}}{1+\left(e^{-y'}+e^{-y}\right){\theta}^{c-1}+ e^{-(y+y^o)}{\theta}^{2c-o-1}} \right \rangle  \\
\frac{o}{N}=\left\langle \frac{e^{-y-y^o}{\theta}^{2c-o-1}}{1+\left(e^{-y'}+e^{-y}\right){\theta}^{c-1}+ e^{-(y+y^o)}{\theta}^{2c-o-1}}\right\rangle.\end{array}\right.
\eea
Using  the asymptotic expansions of these saddle point equations  valid for $c<c^*$ we found
\bea
{\langle{N}^2_c \rangle} & \leq & \sum_{o=0}^c \left( \frac{N}{(c-o)}\left\langle {\theta}^{c-1}\right\rangle \right)^{2(c-o)} \left( \frac{N}{o}\left\langle{\theta}^{2c-o-1}\right \rangle \right)^{o} e^{2c-o} \sqrt{\frac{[2\pi]^3}{c(c-o)o}}\nonumber \\
\eea
Using also for $\langle {N}_L\rangle$ the asymptotic expression $(\ref{asym})$ then we can express the ratio  $\frac{{\langle{N}^2_c \rangle}}{{\langle{N}_c \rangle}^2}$ as
\bea
\frac{{\langle{N}^2_c \rangle}}{{\langle{N}_c \rangle}^2}&\leq&\sum_{o=0}^c\frac{c^{2c}}{(c-o)^{2(c-o)}o^o} \left(\left\langle N{\theta}^{c-1}\right\rangle \right)^{-2o}
\left({N}\left\langle{\theta}^{2c-o-1}\right \rangle \right)^{o}\sqrt{\frac{2\pi c}{o(c-o)}}
\nonumber \\
&\leq  & \sum_{o=0}^c\left(\frac{c^{2c}}{(c-o)^{2(c-o)}o^o e^o}\right) \left(\frac{(c-\gamma)^2}{2c-o-\gamma}\frac{1}{N \left[\langle q\rangle N\right]^{1-\gamma}(1-\epsilon)^{o-\gamma}}\right)^o\sqrt{\frac{2\pi c}{o(c-o)}}
\nonumber \\
&\leq  & \sum_{o=0}^c\left(\frac{c^{2c}}{(c-o)^{2(c-o)}o^o e^o}\right)  \left(\frac{e(c-\gamma) (1-\epsilon)^{(\bar{c}-c)}}{\bar{c}(\bar{c}-\gamma)}\right)^o\sqrt{\frac{2\pi c}{o(c-o)}}.\eea
We notice that  in the limit $c\rightarrow \infty$ we have
\be
\frac{c^c}{e^o(c-o)^{c-o}}\leq c^o\frac{1}{e^o\left(1-\frac{o}{c}\right)^c}\rightarrow c^o
\ee
Using this limit behavior and  Stirling approximations for factorials, we
get
\bea
\frac{{\langle{N}^2_c \rangle}}{{\langle{N}_c \rangle}^2}&\leq \left[1+ \frac{c(c-\gamma) (1-\epsilon)^{(\bar{c}-c)}e}{\bar{c}(\bar{c}-\gamma)}\right]^c
\label{n2_lowerb.eq}
\eea
\begin{itemize}
\item{If $\epsilon=0$}
It is useful to   define the clique size $\hat{c}$ satisfying 
\be 
\frac{\hat{c}(\hat{c}-\gamma)e}{\bar{c}(\bar{c}-\gamma)}=\frac{1}{\hat{c}}
\label{chat.def}
\ee
i.e. $\hat{c}\sim \bar{c}^{2/3}$.
Then if $c=\alpha \hat{c}^{1-\eta}$ we have in the limit $N\rightarrow \infty$, $\bar{c}\rightarrow \infty$,
\bea
\frac{{\langle{N}^2_c \rangle}}{{\langle{N}_c \rangle}^2}&\leq \left\{
\begin{array}{lll} 
1 &\mbox{if} &\eta>0\nonumber \\
e &\mbox{if} &\eta=0\nonumber \\
\infty  &\mbox{if} &\eta<0.
\end{array} \right.
\label{Nub1}
\eea

From Eqs. $(\ref{asym})$ and $(\ref{Nub1})$ for $c=\hat{c}$ defined in $(\ref{chat.def})$ one find that with $c=\alpha \hat{c}^{1-\eta}$
\bea
P({N}_{\hat {c}}>0)\geq \left\{\begin{array}{lll}1 &\mbox{if}& \eta>0 \nonumber \\
\frac{1}{e} & \mbox{if}& \eta=0\nonumber \\
0 & \mbox{if}&\eta<0
\end{array} \right.
\eea
 Consequently  the network contains  almost surely cliques  of sizes $c\leq \underline{c}$ with
\be
\underline{c}=\alpha \hat{c}= \alpha' \bar{c}^{2/3}
\ee
and $\alpha >0$
\item{If $\epsilon>0$}, and  $c= \bar{c}-\alpha \bar{c}^{\eta}$ with $\eta>0$, we have in the limit $N\rightarrow \infty$, $\bar{c}\rightarrow \infty$,
\bea
\frac{{\langle{N}^2_c \rangle}}{{\langle{N}_c \rangle}^2}&< \left\{
\begin{array}{lll} 
1 &\mbox{if} &\eta>0\nonumber \\
\infty  &\mbox{if} &\eta \leq 0.
\end{array}\right.
\label{Nub2}
\eea
From Eqs. $(\ref{asym})$ and $(\ref{Nub2})$ it follows that as long as $ c=\bar{c}-\alpha \bar{c}^{\eta}$
\bea
P({N}_{{c}}>0)\geq \left\{ \begin{array}{lll} 1 &\mbox{if} &\eta>0\nonumber \\ 0 &\mbox{if} &\eta=0,\alpha=0 \end{array} \right.
\eea
Consequently we have that  the network contains almost surely cliques  of sizes $c\leq \underline{c}$ with
\be
\underline{c}=\bar{c}-\alpha \bar{c}^{\eta}
\ee
and $\alpha,\eta >0$
\end{itemize}
 Since if a graph contains a clique of size $\underline{c}$ it contains clearly also  cliques of smaller size we proved that typical networks have a finite probability to get  any cliques of size  $c\leq \underline{c}$.

\begin{table}
\centering
\begin{tabular}{|c|c|c|c|c|}
\hline
\hline
  & \multicolumn{2}{|c|}{$\epsilon=0$} & \multicolumn{2}{|c|}{$\epsilon\neq 0$}\\
\hline
$\gamma>3$ &\multicolumn{4}{|c|}{$c_{max}=3$}\\ \hline
$2<\gamma<3$ &\multicolumn{2}{|c|}{$\underline{c}\leq c_{max}\leq \bar{c}$} &\multicolumn{2}{|c|}{$\underline{c}\leq c_{max}\leq \bar{c}$}\\ 
& \multicolumn{2}{|c|}{} & \multicolumn{2}{|c|}{}\\
& \multicolumn{2}{|c|}{$\bar{c}\sim\bar{c}=N^{\frac{3-\gamma}{4}}$} & \multicolumn{2}{|c|}{$\bar{c}=\frac{3-\gamma}{2}\frac{\log(N)}{|\log(1-\epsilon)|}+O(\log(c))$} \\ 
& \multicolumn{2}{|c|}{} & \multicolumn{2}{|c|}{}\\
 & \multicolumn{2}{|c|}{$\underline{c}=\alpha' \bar{c}^{2/3}$  with $\alpha'>0$} & \multicolumn{2}{|c|}{$\underline{c}=\bar{c}-\alpha \bar{c}^{\eta}$ with $\alpha,\eta>0$}  \\ \hline
$1<\gamma\leq 2$ &\multicolumn{2}{|c|}{$\underline{c}\leq c_{max}\leq \bar{c}$} &\multicolumn{2}{|c|}{$\underline{c}\leq c_{max}\leq \bar{c}$}\\ 
& \multicolumn{2}{|c|}{} & \multicolumn{2}{|c|}{}\\
& \multicolumn{2}{|c|}{$\bar{c}\sim  N^{\frac{1}{2\gamma}} $} & \multicolumn{2}{|c|}{$\bar{c}=\frac{1}{\gamma}\frac{\log(N)}{|\log(1-\epsilon)|}+O (\log(N))$} \\ 
& \multicolumn{2}{|c|}{} & \multicolumn{2}{|c|}{}\\
 & \multicolumn{2}{|c|}{$\underline{c}= \alpha' \bar{c}^{2/3}$ with $\alpha'>0$} & \multicolumn{2}{|c|}{$\underline{c}= \bar{c}-\alpha \bar{c}^{\eta}$ with $\alpha,\eta>0$}  \\\hline
\hline 
\end{tabular}
\caption{}
\label{table}
\end{table}

\subsection{ Average number of cliques passing  through a node}

To find the  expected  number of cliques of size $c$ passing through a given
node, with hidden variable $\bf{q}$, we can repeat the arguments proposed for the calculation of the  first moment with the difference that we integrate over all the hidden variables of the nodes in the cliques except for the hidden variable ${q}_i={\bf{q}}$ of the chosen node. 
Following these arguments one finds for cliques $c<c^*$ 
\be {N}_c({\bf{q}})\simeq \left(\frac{{\bf q}}{\sqrt{\langle q\rangle N}}\right)^{c-1} {N}_{c-1}.
\ee
Consequently nodes with higher hidden variable $\bf{q}$ are expected to be part of more cliques.

\section{Molloy Reed ensembles}

The counting of the number of cliques in the Molloy-Reed \citep{MR}
follows a procedure much similar to the one considered for the hidden variable ensemble giving similar results. 
To construct a Molloy-Reed network one proceed as follows:
{\it i)} a degree is assigned to each node of the network following
the desired degree distribution with cutoff $K$
\bea
K\sim \left\{ \begin{array}{lcr}
N^{1/\gamma} &\mbox{for} & \gamma \in (1,2]\nonumber \\
N^{1/2} & \mbox{for}& \gamma \in (2,3]\nonumber \\
N^{\frac{1}{\gamma-1}} & \mbox{for}& \gamma\in (3,\infty) 
\end{array}\right ..
\eea 
Degree distributions which do not
satisfy the parity of $\langle k \rangle N=\sum_i k_i$ are disregarded; 
{\it ii)} the edges  coming out of the nodes are randomly matched
until all edges are connected.  The structural cutoff for $\gamma\leq 3$ ensures that the probability of double links and tadpoles is small \cite{lungo}.

To calculate $\langle {N}_c \rangle$ in this ensemble first one has to count in
how many ways it is possible to have a clique of size $c$ in the network
and weight the results with the fraction of possible networks in the
ensemble which  contains the clique. 
Let us first state that the total number of graphs in the Molloy-Reed
ensemble is given by $(\langle k\rangle N-1)!!$. Indeed when constructing the network
by linking $\langle k \rangle N$ unconnected edges one start by taking one edge at
random and connecting it to one of the $(\langle k \rangle N-1)$ possible
connections. Then one proceed taking another edge and linking it to
one of the remaining  $(\langle k \rangle N-3)$ possible connections thus giving rise
of one of the $(\langle k \rangle N-1)!!$ possible networks. 
By similar arguments one shows that the total number of networks
containing a given clique of size $c$ are $[\langle k \rangle N-c(c-1)-1]!!$. 
On the other side the total number of cliques of size $c$ in the
Molloy-Reed ensemble is given by the number of ways one can choose $c$ nodes  $\{1_i,\ldots,i_c\}$  of connectivity
$\{k_1,k_2,\dots,k_c\}$ and connect each pair of them. 
The number of ways one can choose the edges coming out of the nodes to form the clique is given by
\[
\Pi_{i=1}^c \frac{k_i!}{(k_i-c+1)!}.
\]
Consequently the average number of cliques in the Molloy-Reed ensemble will be given by
\be
{N}_c= \sum_{\{n_k\}}  \prod_{k=c}^K  \left(\begin{array}{l} N(k)\\ n(k) \end{array} \right) \left(\frac{k!}{(k-c+1)!}\right)^{n_k}W_{N,c}
\label{unoMR.eq}
\ee
where $N(k)=NP(k)$ ($n(k)$) is the number of nodes with connectivity $k$
present in the network (loop), $K$ is the  cutoff of the degree distribution and  the sum over $\{n(k)\}$ is restricted to  $\{n(k)\}$
such that $\sum_k n(k)=c$. Moreover we use the definition $W_{N,c}={(\langle k\rangle N-c(c-1)-1)!!}/{(\langle k \rangle N-1)!!}$.
If we use the Stirling approximation for  $W_{N,c}$ 
we get the expression
\be
 W_{N,c}\sim (\langle k\rangle N)^{-c(c-1)/2}  e^{Ng(\omega)}
\ee
with $\omega=c(c-1)/N$ and
\be
g(\omega)=\frac{1}{2}(\langle k\rangle-2\omega)\log\left(\frac{\langle k\rangle-2\omega}{\langle k\rangle}\right)+\omega\sim \frac{3\omega^2}{\langle k\rangle}
\label{g.eq}
\ee
Thus we get
\be
{N}_c= \sum_{\{n_k\}}  \prod_{k=c-1}^K \left(\begin{array}{l} N(k)\\ n(k) \end{array} \right)  \left({\kappa}^{c-1}\right)^{n(k)}e^{N g(\omega)}
\label{MRc}
\ee
where 
\be
{\kappa}^{c-1}=\frac{k!}{(k-c+1)!}\frac{1}{(\langle k \rangle N)^{(c-1)/2}}
\ee
Expression $(\ref{MRc})$  for the average number of cliques in a Molloy Reed ensemble differs from the equivalent expression in the hidden variable ensemble  $\ref{N_m}$ {\em i)} for the substitution ${\theta}^{c-1}\rightarrow {\kappa}^{c-1}$; {\em ii)} for the factor $\exp(Ng(\omega))$ and {\em iii)} for the fact that the average is performed only on the nodes with connectivity $k\geq c-1$.
Following the same steps as in the hidden variable  ensemble, we get 
\be
{N}_c=
\int_{-\infty}^\infty \frac{dy}{2\pi} e^{cy+N\left\langle \log\left[1+{\kappa}^{c-1} e^{-y}\right]\right\rangle_{c-1}+N g(\omega)}
\label{exactMR.eq}
\ee 
with  $g(\omega)$ given by Eq. (\ref{g.eq}) and the average performed of the $N(k) $ distribution with a lower cutoff at $k=c-1$.

Evaluating $(\ref{exactMR.eq})$ by the saddle point method and following the steps described in the preceding section,  we get the following approximate expression for the average number of cliques $N_c$
\begin{equation}\label{NsmallMR.eq}
  {N}_c=\left(\frac{e \langle {\kappa}^{c-1}\rangle_{c-1}}{c}\right )^c
\end{equation}
where this approximation is valid asymptotically for cliques of  sizes $c<c^*$
We note that $c^*$ is fixed by the condition
\be
\frac{c}{N}
\frac{  \left\langle\left({\kappa^{c-1}}\right)^{2} \right \rangle }{
\left[\left\langle{\kappa}^{(c-1)}\right \rangle \right]^2}\leq K^{\gamma-1}\frac{c^2}{(2c-3)N} < 1.
\label{bounds2.eq}
\ee
Similar results to the one found for the hidden-variable ensemble also apply for the second-moment of the number of cliques in the Molloy-Reed ensemble.

\section { Conclusions}
In conclusion we have  have calculated the first and the second moment of the number of cliques in random scale-free network ensembles.
 This calculation show  these networks, provided that the power-law exponent $\gamma<3$ have many small cliques and a large clique number.
In particular the clique number diverges with the network size as long as $\gamma<3$ which is a surprising results since in Erd\"os and Renyi random networks with finite average degree the maximal clique size is $c_max=3$. Moreover we have shown that in the case in which the cutoff is the maximal allowed cutoff (i.e. following the terminology of the paper when  $\epsilon=0$) there can be large fluctuations of the clique number wherever for $\epsilon \neq 0$ the fluctuations are small.

\appendix
\section{ Calculation of the upper bound for the clique number in the case $\epsilon=0$ in the hidden variable ensemble}

The evaluation of the upper bound for the clique number in the subtle case $\epsilon=0$ deserve a particular attention.
To address this problem we start by   rewriting  in the following the main results for the average number of cliques in the hidden variable ensemble. The expression $(\ref{N_msp})$ for the average number of cliques is given by
\be
\langle N_c\rangle=  e^{N f(y^*) }\sqrt{\frac{2\pi}{N f''(y^*)}}
\label{N_msp2}
\ee
where $y^*$ is provided by the saddle point equation $(\ref{ystar})$
\be
\frac{c}{N}=\left\langle\frac{{\theta}^{c-1} e^{-y^*}}{1+{\theta}^{c-1} e^{-y^*}}\right\rangle.
\label{sp.eq_a}
\ee
and  $f(y^*)=y^*c+N\langle\log [1+{\theta}^{c-1} e^{-y^*}] $ while  $\theta$ is given by
 \be
{\theta}=\left(\frac{q}{\sqrt{\langle{q}\rangle N}}\right).
\ee
If $y^*>0$ then we have that 
\be
y^*c+N \langle \log [1+{\theta}^{c-1} e^{-y^*}]\rangle \leq y^* c+N \langle \log [1+{\theta}^{c-1}]\rangle.
\label{uno_a}
\ee
On one side, from  the saddle point equation Eq. $(\ref{sp.eq_a})$ we have
\be
e^{y*}\leq  {\left\langle\frac{N \theta^{c-1}}{c}\right\rangle}^c.
\label{due_a}
\ee
on  the other side we have that,
\bea
N \langle \log [1+{\theta}^{c-1}]\rangle\leq 2 N \left\langle\frac{{\theta}^{c-1} e^{-y^*}}{1+{\theta}^{c-1} e^{-y^*}}\right\rangle= 2c.
\label{tre_a}
\eea
Moreover  the second derivative $f''(y^*)$ satisfy
\bea
N f''(y^*)&=&N\left\langle \frac{{\theta}^{c-1} e^{-y*}}{(1+e^{-y*}{\theta}^{c-1})^2}\right \rangle \nonumber \\
&\geq &  N\left\langle \frac{e^{-y*}{\theta}^{c-1}}{(1+e^{-y*} {\theta}^{c-1})}\right \rangle\frac{1}{1+e^{-y^*}}\nonumber \\
&\geq &\frac{c}{{2}}. 
\label{qu_a}
\eea
Consequently, putting together Eqs. $(\ref{uno_a})$, $(\ref{due_a})$ $(\ref{tre_a})$ and finally Eq. $(\ref{qu_a})$ the average number of cliques $\langle { N}_c \rangle$  $(\ref{N_msp2})$ satisfy
\be
\langle {N}_c \rangle \leq \sqrt{\frac{2}{c}} {\left\langle\frac{N \theta^{c-1}\e^2}{c}\right\rangle}^c.
\ee 
which together with the inequality $(\ref{undici})$ provides the upper bound $(\ref{cbar})$ for the clique number scales with the system size as shown in Table $\ref{table}$.

At this point we must check self-consistently that indeed is $c<\bar{c}$ then $y^*>0$. To prove this we suppose on the contrary that $y^*<0$. In this eventuality, the saddle point equation can be rewritten as
\bea
\frac{c}{N}&=& \sum_{q<\bar{q}} \rho(q) \frac{{\theta}^{c-1} e^{-y^*}}{1+{\theta}^{c-1} e^{-y^*}} + \sum_{q>\bar{q}} \rho(q)\frac{{\theta}^{c-1} e^{-y^*}}{1+{\theta}^{c-1} e^{-y^*}}
\eea
where $\bar{q}=Q e^{\frac{y^*}{(c-1)}}<Q$.
Expanding the two terms in series we get 
\be
\frac{c}{N}=e^{-y^*\frac{\gamma-1}{c-1}} N \frac{Q^{1-\gamma}}{c} (F_{\gamma,c}+G_{\gamma,c})
\ee 
with 
\bea
F_{\gamma,c}&=&m^{\gamma-1}(\gamma-1) \sum_{n=1}^{\infty} (-1)^{n}\frac{c}{[(c-1)n+1-\gamma]}\rightarrow F^*_{\gamma}\nonumber \\
G_{\gamma,c}&=&m^{\gamma-1}(\gamma-1) \sum_n (-1)^n\frac{c}{[(c-1)n+\gamma-1]}\rightarrow G^*_{\gamma}.
\eea
Therefore for $c\gg 1$ we have
\be
e^{-y^*}=\left(\frac{c^2}{N (F_{\gamma}^*+G_{\gamma}^*)^2}\right)^{\frac{c-1}{\gamma-1}}
\ee
Lets observe that  $\bar{c}$ in given by the value in the table $\ref{table}$ always satisfy  $\bar{c} \ll N^{1/2}$ for $\gamma>1$. Moreover as long as $c\rightarrow \infty $  with $c \ll N^{1/2}$, we get form expression $(\ref{ysf})$ that $y^*\rightarrow 0^+$.
Consequently we assuming $y^*>0$ for $c>\bar{c}$ we have reached a contradiction. 
This proves that in the hypothesis $c>\bar{c}$ the saddle point solution  $y^*$ is always positive, i.e. $y^*>0$ as we assume at the beginning of the paragraph.

\end{document}